\documentclass{article}

\usepackage{PRIMEarxiv}

\usepackage[utf8]{inputenc} 
\usepackage[T1]{fontenc}    
\usepackage{hyperref}       
\usepackage{url}            
\usepackage{booktabs}       
\usepackage{amsfonts}       
\usepackage{amsmath}
\usepackage{nicefrac}       
\usepackage{microtype}      
\usepackage{lipsum}
\usepackage{fancyhdr}       
\usepackage{graphicx}       
\usepackage{natbib}
\usepackage{eurosym}
\usepackage{booktabs,tabularx}
\usepackage{rotating}
\usepackage{pdflscape}
\usepackage{graphicx}
\usepackage{subcaption}
\usepackage{placeins}        
\usepackage{svg}
\usepackage{multirow}

\usepackage{algorithm}
\usepackage{algpseudocode}

\usepackage{todonotes}
\let\opentask\todo 
\renewcommand{\todo}[1]{\opentask[inline,color=red!40]{#1}}

\title{Effects of Dynamic and Stochastic Travel Times on the Operation of Mobility-on-Demand Services
}

\author{
  Fynn Wolf, Roman Engelhardt, Yunfei Zhang, Florian Dandl and Klaus Bogenberger\\
  Chair of Traffic Engineering and Control \\
  Technical University of Munich \\
  Arcisstraße 21 \\
  80333 Munich, Germany \\
  \texttt{Corresponding Author: roman.engelhardt@tum.de} \\
}

\begin{document}
\maketitle

\begin{abstract}
Mobility-on-Demand (MoD) services have been an active research topic in recent years. Many studies focused on developing control algorithms to supply efficient services. To cope with a large search space to solve the underlying vehicle routing problem, studies usually apply hard time-constraints on pick-up and drop-off while considering static network travel times to reduce computational time. As travel times in real street networks are dynamic and stochastic, assigned routes considered feasible by the control algorithm in one time step might become infeasible in the next. Since once assigned and confirmed, customers should still be part of the solution, damage control is necessary to counteract this effect. In this study, a detailed simulation framework for MoD services is coupled with a microscopic traffic simulation to create dynamic and stochastic travel times, and tested in a case study for Munich, Germany. Results showed that the combination of inaccurate travel time estimation and damage control strategies for infeasible routes deteriorates the performance of MoD services -- hailing and pooling -- significantly. Moreover, customers suffer from unreliable pick-up time and travel time estimations. Allowing re-assignments of initial vehicle schedules according to updated system states helps to restore system efficiency and reliability, but only to a minor extent.
\end{abstract}


\section{Introduction}

Mobility-on-Demand (MoD) services started to play a vital role in urban mobility in recent years. Especially considering upcoming autonomous vehicles (AVs), large-scale services with cheap fares can be offered. The introduction of such convenient service might decrease parking space consumption by replacing private vehicles trips and additionally vehicle miles traveled if trips are shared in pooled services. 

Travel time estimation and prediction is an important aspect for MoD services. From an operator's point of view, vehicle routing and dispatching heavily relies on travel time estimations to evaluate costs of possible routing decisions and check the feasibility of schedules regarding customers' pick-up and drop-off time constraints. From a customer's point of view, a convenient service should provide accurate information about expected pick-up and drop-off times, in the best case even before they book the service. Reliability and convenience of a service will suffer heavily if those estimations are incorrect, especially for long-term acceptance of shared autonomous vehicle services~\citep{narayanan2020shared}.


Many recent studies focused on developing operational strategies for hailing (e.g. \citep{Zhang.2016, hyland2018dynamic} ...) or pooling services (e.g. \citep{AlonsoMora.2017, Engelhardt.1027201910302019, Kucharski.2020} ...). Nevertheless, usually deterministic network travel times from historic or simulated data is applied. On the one hand, this assumption allows the preprocessing of routing tables to reduce the need to compute fastest routes, which otherwise becomes the computational bottlenecks of solving the assignment problem. On the other hand, hard time constraints on customer pick-up and drop-off are usually applied to further reduce the search space of the underlying vehicle routing problem. If network travel times are deterministic and known, it can be guaranteed that a feasible vehicle schedule assigned once will also remain feasible in the future \citep{Dandl.2021}.

These models therefore lack at evaluating the traffic effect the MoD service has on the overall system and additionally, they fail in providing damage control strategies to deal with infeasible solutions when travel time estimations are flawed.


To evaluate the traffic impact of MoD services, different methods are developed to achieve the balance between accuracy and computational efficiency. They vary from retrieving dynamics from map services~\citep{markov2021simulation}, link transmission models~\citep{levin2017congestion}, or direct coupling with a microscopic simulation such as Aimsun Next~\citep{dandl2017microsimulation}. To reduce the need of fastest path computations, macroscopic fundamental diagram (MFD) based methods have been proposed to estimate traffic and travel times in different network regions~\citep{dandl2021regulating}.

All these studies have in common, that they utilize complex traffic models to investigate the impacts of MoD services on the traffic system itself. Nevertheless, the operational implications of arrival reliability are not evaluated. This reliability issue can be degraded into two sub-problems, namely 1) uncertainty regarding waiting and traveling times and 2) route changes due to stochastic requests.
To the authors' knowledge, only~\cite{liu2019dynamic} deals with the uncertainty regarding waiting and traveling times of MoD services. They compared different path finding algorithms (reliable or shortest) and different information (historical or real-time). The results shown that using the most reliable path model (with the largest probability of arrival within a specified time budget) and historical travel time can significantly improve the on-time arrival reliability. \cite{fielbaum2020unreliability} systematically analyzes the general uncertainty for MoD services, which can be caused by other users or by the operational rules.

With respect to the provided literature, we seek to answer the following two research questions:

\begin{enumerate}
    \item How do dynamic and stochastic travel times affect the fleet performance indicators of MoD services?
    \item How do dynamic and stochastic travel times affect the difference between expected and realized waiting time?
\end{enumerate}

To answer these two questions, we evaluate the effects of dynamic travel times on a MoD services by coupling the MoD simulation framework "FleetPy"\footnote{https://github.com/TUM-VT/FleetPy}~\citep{Engelhardt.28.07.2022} with the microscopic traffic simulator Aimsun Next and compare the effects of dynamic and stochastic travel times to the often made assumption of static and deterministic travel times.


Our contributions can be summarized as follows:

\begin{itemize}
    \item Coupling of a MoD simulation framework with microscopic traffic simulation
    \item Developing a damage control strategy to cope with stochastic travel times
    \item Evaluation of the impact of dynamic and stochastic travel times on users' and operators' indicators 
    \item Evaluation of the impact of re-assignments with dynamic and stochastic travel time
\end{itemize}

\section{Methodology}

\begin{figure*}[t]
    \centering
    \includegraphics[width=1.0\columnwidth]{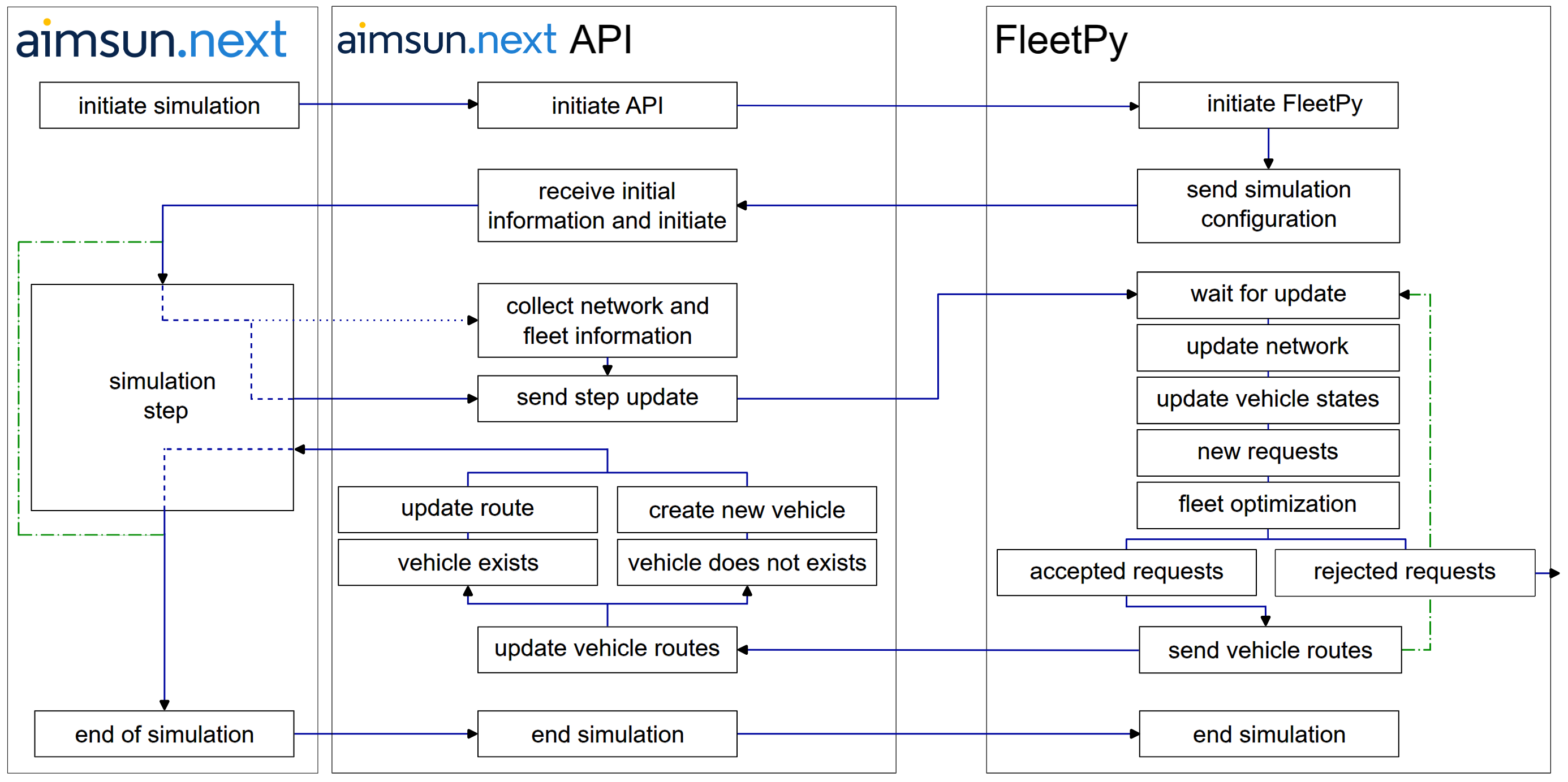}
    \caption{High level flowchart of the coupled simulation of Aimsun Next and FleetPy.}
    \label{fig:workflow_FleetPy_Aimsun}
\end{figure*}

First, the fleet simulation tool FleetPy is described. Inherently, FleetPy focuses on the control aspects and uses a simplified traffic representation with deterministic travel times. To evaluate the impacts of this simplification, in the second part of this section the coupling of FleetPy with the microscopic traffic simulator "Aimsun Next"\footnote{https://www.aimsun.com/aimsun-next/} is described. With this coupling, FleetPy controls the fleet, i.e. assigns vehicle schedules and provides feedback to customers, while vehicle movement is performed in the microscopic traffic simulator.  

\subsection{FleetPy}
\label{chap:FleetPy}
FleetPy is an agent-based simulation model to analyze MoD fleets. For this study, three main types of agents are of relevance: 1) Customers requesting trips from the MoD service, 2) an operator that offers the service by (centrally) controlling a set of vehicles, and 3) the fleet vehicles picking up and dropping off customers according to the schedules assigned by the operator.

To estimate vehicle travel times, the operator represents the street network as a directed graph $G_{op} = (N_{op}, E_{op})$ with nodes $N_{op}$ and edges $E_{op}$. Each edge $e \in E_{op}$ is associated with a distance $d_e$ and a travel time $\tau_e (t)$. Depending on the simulation time $t$ and the existing traffic state, a goal of the operator is to estimate edge travel times $\tau_e (t)$ to accurately predict times for vehicles to arrive at scheduled stops.

Four main steps are conducted during a FleetPy time step:
\begin{enumerate}
    \item Boarding and alighting processes of vehicles are registered by the fleet operator.
    \item New MoD customers enter the simulation and request a trip $i$ at time $t_i$ by providing origin $o_i \in N_{op}$ and destination $d_i \in N_{op}$.
    \item The operator evaluates whether it can serve the request within the given time constraints. If so, an expected pick-up time $t_i^{pu}$ and drop-off time $t_i^{do}$ is provided.
    \item The operator assigns new/updated schedules to its vehicles.
\end{enumerate}

Assigning new schedules is one core feature of FleetPy and will be described on a high level in the following. A schedule is defined as a series of stops at network nodes $N_{op}$ where boarding and alighting processes of vehicles are conducted. In between these stops, vehicles are traveling on the fastest route in the network $G_{op}$. There are multiple possible permutations of stops as soon as more than one passenger is assigned to a vehicle $v \in V$, which are enumerated. The $k$-th possible permutation of stops for the schedule $\psi_k(v, R_\psi)$ serving all requests in the set $R_\psi$ is considered feasible if
\begin{enumerate}
    \item the drop-off stop succeeds the pick-up stop for each customer.
    \item the number of on-board customers never exceeds the vehicle capacity ($c_v$).
    \item each customer is (supposed to be) picked up before a maximum waiting time $w_{max}$ elapsed.
    \item if the operator offers a pooling service, the maximum additional travel time must not exceed a detour factor $\delta_{max}$ compared to a direct trip.
\end{enumerate}

Each feasible schedule is rated by a cost function to be minimized, defined as

\begin{align}
    \nonumber
    \rho(&\psi_k(v, R_\psi)) = t_{end} (\psi_k) - t_{sim} - P_r |R_\psi| + \\
    \nonumber
    &+ P_{delay} \sum_{r \in R_\psi} (\text{max}[t_{r,pu}(\psi_k) - t_{r,pu}^{latest}, 0] ) + \\
    &+ P_{delay} \sum_{r \in R_\psi} (\text{max}[t_{r,do}(\psi_k) - t_{do}^{latest}, 0] ) ~.
\end{align}
$t_{end} (\psi_k)$ refers to the expected time this schedule is finished. Together with the current simulation time $t_{sim}$, the first term measures the time of execution. With the number of served requests by the schedule $|R_\psi|$ and a large assignment reward $P_r = 1.000.000$s, the objective prioritizes serving as many customers as possible. The last two terms penalize delays in arrival times of the vehicle. $t_{r,pu}(\psi_k)$ and $t_{r,do}(\psi_k)$ refer to pick-up and drop-off times of the requests according to the plan, respectively. $t_{r,pu}^{latest}$ and $t_{do}^{latest}$ reflect their latest pick-up and drop-off times. $P_{delay} = 10$ weighs the delays if present. Conflating all possible permutations, the cost $\rho_\psi$ of serving the set of requests $R_\psi$ with vehicle $v$ is defined as that permutation $k$ of $\psi_k(v, R_\psi)$ that minimizes $\rho(\psi_k(v, R_\psi))$.

At each optimization interval, the operator assigns new feasible schedules to accommodate new customers by solving the following integer linear problem:

\begin{align}
    \text{minimize} & \sum_{v \in V,j \in \Psi} \rho_{vj} x_{vj} & \\
    \label{eq:const1}
    \text{s.t.} & \sum_{j \in \Psi} x_{vj} \leq 1 & \forall v \in V \\
    \label{eq:const2}
    & \sum_{v \in V, j \in \Psi_r} x_{vj} = 1 & \forall r \in R_a \\
    \label{eq:const3}
    & \sum_{v \in V, j \in \Psi_r} x_{vj} \leq 1 & \forall u \in R_u
\end{align}
$x_{vj} \in \{0, 1\}$ is a binary decision variable to assign a schedule $j$ of the set of all available schedules $\Psi$ to vehicle $v$. Eq.~\eqref{eq:const1} ensures that not more than one schedule is assigned to each vehicle. Eq.~\eqref{eq:const2} ensures that each request from the set of requests that have been assigned in previous time step $R_a$ are assigned again. Contrarily, yet unassigned requests from the set $R_u$ can remain unassigned as stated in constraint~\eqref{eq:const3}. Thereby, $\Psi_r$ refers to the set of schedules that serve request $r$.

If an unassigned request $r \in R_u$ is not assigned to a vehicle, it is assumed that this request leaves the system unserved. On the other hand, if it is assigned, the corresponding assigned schedule is used to determine the expected pick-up $t_{i, pu}^{exp}$ and travel time which are provided to the customer. Additionally, the request is moved to the set $R_a$ in the next optimization step.

The core of the algorithm is to create the set of feasible schedules $\Psi$. The algorithm applied in this study is a variant of the constraint programming approach developed by~\cite{AlonsoMora.2017}. By explicitly exploiting the time constraint for feasible schedules, a systematic exhaustive search can be done to create all feasible routes. Because of constraints in the length of this paper, we refer to original publication. Details of the implementation can be found in~\cite{Engelhardt.29.07.2020}.

This study distinguishes two options of set of schedules that are created for the assignment problem:
\begin{enumerate}
    \item No Re-assignment: Once a request $r$ is assigned to vehicle $v$, no schedule of another vehicle to serve request $r$ is included in the assignment problem to guarantee that a request remains assigned to the same vehicle. This might be convenient for the customer because the assigned vehicle position can directly be provided and its approach can be tracked. However, optimization potential might be lost.
    \item With Re-assignment: Here, every feasible schedule is included in the assignment problem and the assigned vehicle for each customer might change while still fulfilling pick-up and drop-off constraints. This re-assignment occurs when a better global assignment solution is found.
\end{enumerate}

Due to the coupling with Aimsun, the actual vehicle travel times are not the same as the ones the algorithm for creating feasible schedules assumes. Hence, created schedules that were considered feasible in the last time step might become infeasible due to delays (e.g. congestion or unfortunate signal timings). If only feasible schedules are considered in the assignment problem, the problem --- more specifically equation~\eqref{eq:const2} --- cannot be feasible as soon as $\Psi_r = \empty$ for any previously assigned request $r \in R_a$. Therefore, each currently assigned vehicle schedule is included in the assignment problem, even though it might have become infeasible. In the worst case, the strategy is therefore to stick with the current schedule. It should be noted, that if an assigned vehicle schedule became infeasible, no additional customers can be accommodated until it finished this schedule. In case re-assignments are possible, the cost function penalizes infeasible schedules and closer vehicles could be re-assigned to serve customers in time.

At the end of the assignment stage, idle vehicles are rebalanced to provide supply in the whole network. The algorithm described in \cite{AlonsoMora.2017} is used, which sends idle vehicles to all network locations where customers had to be rejected.

\subsection{Coupling with Aimsun Next}

A simplified overview of the simulation flow of the coupling wit Aimsun Next is shown in Fig.~\ref{fig:workflow_FleetPy_Aimsun}. The simulation flow is separated into three main entities: The Aimsun Next microsimulation, its Aimsun Next API accessible with Python and the fleet control tool FleetPy. On a high level, the simulation time is controlled in the Aimsun Next microsimulation and triggers updates of simulation time and vehicle states in FleetPy via its Python API. FleetPy computes updated vehicle schedules to be performed in Aimsun Next as feedback.

\paragraph*{Network Preprocessing} Aimsun Next and FleetPy are not accessing the same network object. Therefore, network representations have to be translated to allow seamless integration. On a high level, the Aimsun Next network consists of sections and turns that connect different sections within intersections. Each section and turn is converted to a FleetPy edge and added to the network $G_{op}$. The corresponding Aimsun Next object identifier is assigned to each edge to allow the translation between the different networks.

\paragraph*{Simulation Initialization} To begin, the traffic microsimulation and the fleet control are started and initialized. This initialization firstly synchronizes the absolute simulation start time between the coupled frameworks. Additionally, interaction periods are defined. Traffic microsimulations usually have simulation time steps on the sub-second scale. However, the fleet control does not require updates this frequently. Instead, fleet control updates are only triggered every $\Delta t_{fc}$ (in the range of 10 seconds to a minute). Additionally, the frequency of updating network travel times $t_{statistics}$ for the fleet operator is defined. 

\paragraph*{Simulation Flow} At the beginning of each simulation time step, the microsimulation performs vehicle movements. The API tracks fleet vehicles, i.e. their location and if they reached their destination. Boarding processes are controlled in FleetPy. When a vehicle reaches a destination, FleetPy checks if a boarding process has to be started. If so, the actual boarding and alighting times $t_{i, pu}^{act}$ and $t_{i, do}^{act}$ are set, respectively. Likewise, FleetPy checks if boarding process ended and new vehicle routes should start. If so, corresponding routes are sent to the API to create new vehicles in Aimsun. Additionally, every $\Delta t_{fc}$, a new fleet optimization is triggered. First, the positions of vehicles that are currently on-route are set, and their assigned schedule is checked for feasibility. Next, new customer requests for the time step are gathered, and the fleet optimization described in the previous section is triggered. Unassigned requests will leave the system (and look for another mobility option). New vehicle schedules to accommodate accepted customers are assigned to the vehicles. Finally, new routes (i.e. the fastest paths in network $G_{op}$) for vehicles that received a new or updated routing task are collected and returned to Aimsun Next.

To calculate new edge travel times for the fleet operator, the duration vehicles need to pass each Aimsun Next section is tracked within each statistics interval $t_{statistics}$. In this study, not only fleet vehicles, but all vehicles within the microsimulation are tracked. At the end of an interval, the average duration on each edge is provided to the fleet operator as an estimate for the current travel times on each edge.

\section{Case study}

\begin{figure}
    \centering
    \includegraphics[width=0.7\columnwidth]{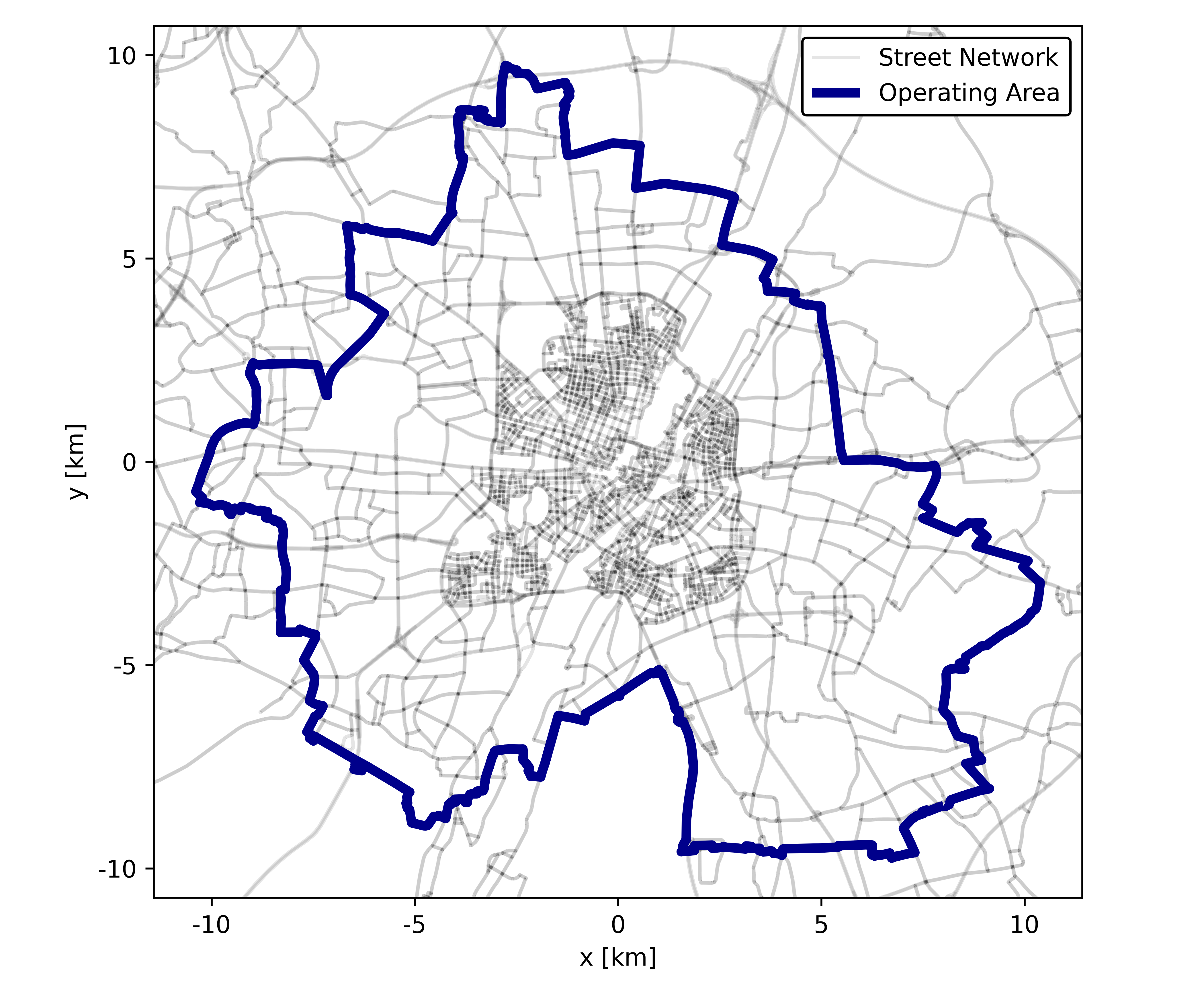}
    \caption{Street network of the Aimsun Next model of Munich.}
    \label{fig:sim_network}
\end{figure}

To evaluate the impact of the coupling of the fleet control tool FleetPy with the microsimulation Aimsun Next, a case study for Munich, Germany is performed. 


\subsection{Aimsun Next Model}

The Aimsun Next model described in \cite{F.Dandl.2017} is used in this study. The network is shown in Fig.~\ref{fig:sim_network}. The operating area of the MoD service is given by the blue boundary, resembling the urban area of the city of Munich.

The model is calibrated using hourly time-dependent private vehicle OD-Matrices and loop detector data for given sections. 
The simulation is run for the morning peak from 6:00 to 10:00 a.m.. In total there are 620k vehicle trips in the microsimulation.

The MoD demand is generated from the private vehicle trip OD matrices. 5\% of those entries that start and end within the operating area shown in Fig.~\ref{fig:sim_network} are used to generate MoD customers with Poisson processes, resulting in 13k trips within the simulation period.

The first and the last $3600$s of each simulation are used as warm-up and cool-down period, respectively, and not included in the evaluation.

\subsection{Scenarios}

In all scenarios, the MoD operator applies a maximum waiting time constraint $w_{max} = 480$s and a maximum detour constraint of $\delta_{max} = 40\%$ is used. The boarding time is assumed to be $30$s. New customers are assigned and new vehicle routes are computed every $\Delta t_{fc}=60$s. Edge travel times in the operator network are updated every $t_{statistics} = 1800$s.

In total, 6 scenarios are created. They differ in at least one of the following attributes:
\begin{enumerate}
    \item \textbf{Coupled/Not Coupled (C/nC):} Coupled scenarios correspond to simulations with vehicle movement performed in Aimsun Next. In not Coupled scenarios, vehicle movements are performed in a deterministic network. To have comparable scenarios, the travel times are extracted from the corresponding coupled scenarios and used as edge travel times in these scenarios.
    \item \textbf{Hailing/Pooling (H/P):} In the hailing service, trips are not shared. Within the simulation scenarios, this is achieved by setting the vehicle capacity to $c_v^{hail} = 1$, while the capacity for pooling vehicles is $c_v^{pool} = 4$. To supply a similar service level, 1550 vehicles are applied for the hailing service and 1150 for the pooling service.
    \item \textbf{With/Without Re-assignment (wR/nR)}: When re-assignment is not allowed, the initially assigned vehicle has to pick up the customer. If re-assignment is allowed, the optimization can change the vehicle assigned to a customer.
\end{enumerate}

\section{Results}


\begin{table*}[t!]
    \centering
    \caption{KPIs of simulated scenarios.}
    \label{tab:eval_scenarios}
    \fontsize{7}{9}\selectfont
        \begin{tabular}{c|c||c|c|c|c|c|c|c|c|c}
                               &         & Served    &             & Avg.      & Saved    & Fleet       & Avg.       & Avg.        & Avg. (Std.)        & Avg. (Std.) \\
            Scenario           & Coupled & Customers & Fleet KM    & Occupancy & Distance & Utilization & Waiting Time  & Travel Time & $\Delta^{wt}$ & $\Delta^{tt}$ \\
                               &         & [\%]      & [$10^3$ km] &    [Per/km]       & [\%]     & [\%]        & [s]        & [s]         & [s]                & [s] \\
            \hline
            \hline
            \multirow{2}{*}{H:nR} & nC     & 94  & 34  & 0.8  & -25 & 54 & 264 & 667  & 2 (24)    & 17 (111)   \\ 
                                 & C     & 94  & 36  & 0.8  & -37 & 72 & 367 & 875  & 106 (270) & 241 (417)   \\ \hline 
            \multirow{2}{*}{P:nR} & nC     & 97  & 35  & 1.2  & 7   & 59 & 253 & 809  & 3 (27)    & 82 (151)  \\ 
                                 & C     & 93  & 40  & 1.0  & -11 & 81 & 396 & 1097 & 124 (339) & 393 (613)  \\ \hline 
            \multirow{2}{*}{P:wR} & nC     & 98  & 35  & 1.2  & 10  & 57 & 255 & 804  & 2 (72)    & 71 (155)  \\ 
                                 & C     & 96  & 40  & 1.1  & -8  & 73 & 342 & 1015 & 60 (209)  & 306 (465) \\ 
        \end{tabular}
\end{table*}

\begin{figure*}[t]
    \centering
    \begin{subfigure}{.33\textwidth}
        \centering
        \includegraphics[width=\linewidth]{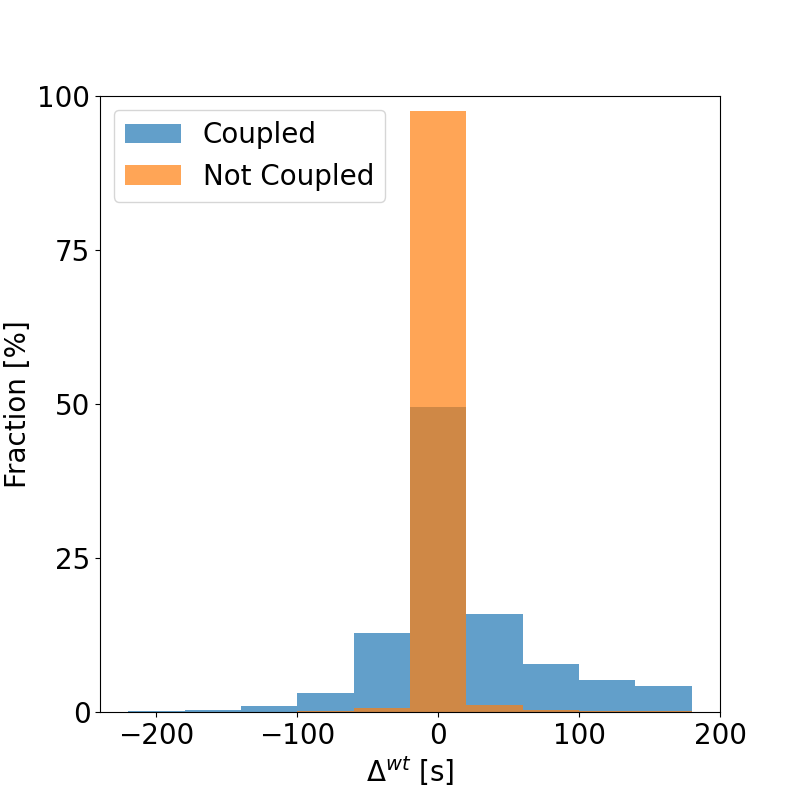}
        \caption{Hailing: No Reassignment (H:nR)}
        \label{fig:hist_hail}
    \end{subfigure}%
    \begin{subfigure}{.33\textwidth}
        \centering
        \includegraphics[width=\linewidth]{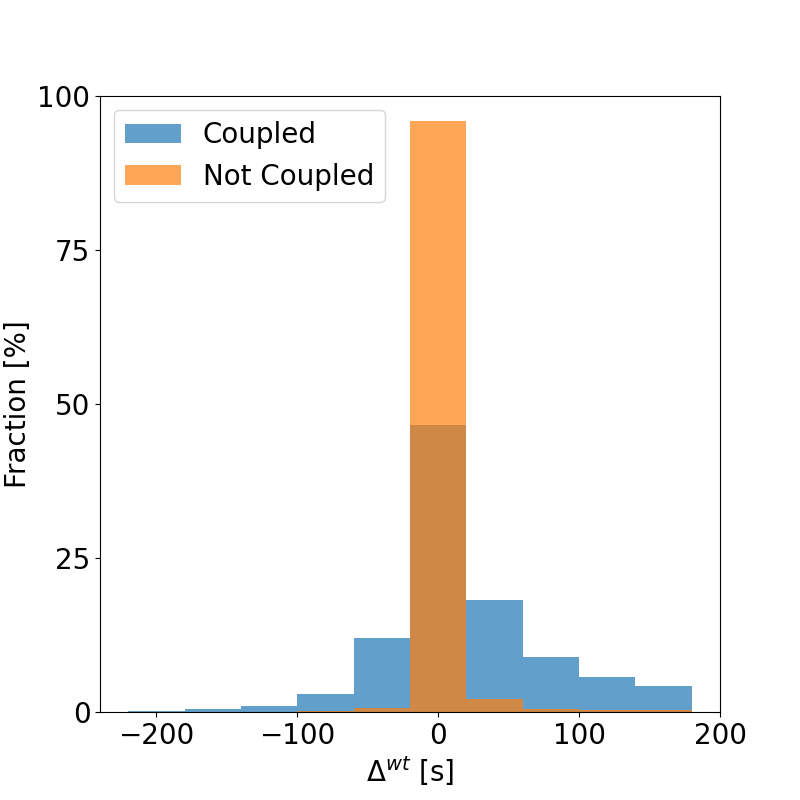}
        \caption{Pooling: No Reassignment (P:nR)}
        \label{fig:hist_pool_no_re}
    \end{subfigure}
    \begin{subfigure}{.33\textwidth}
        \centering
        \includegraphics[width=\linewidth]{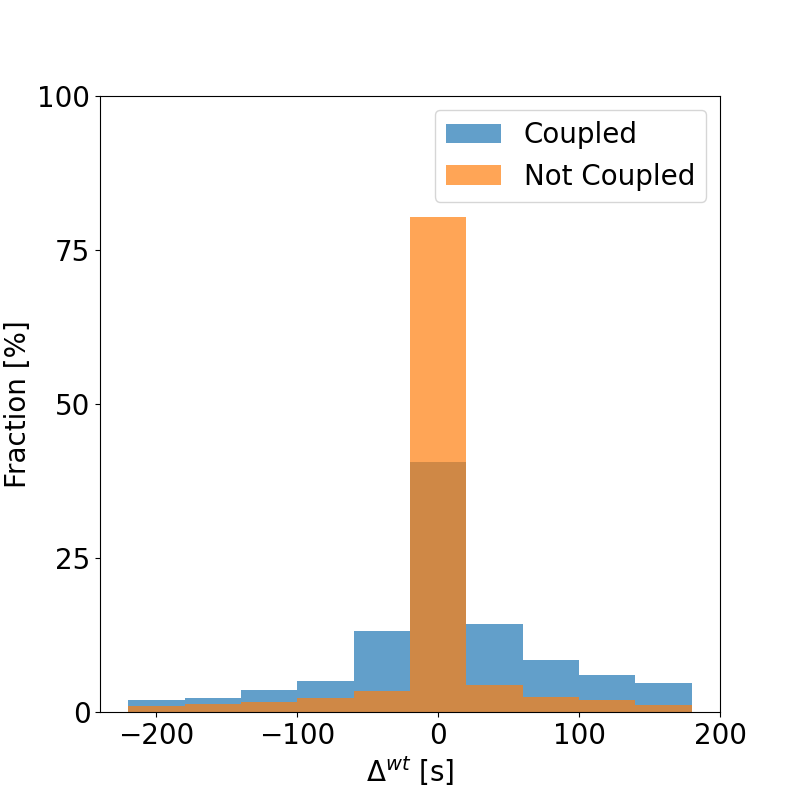}
        \caption{Pooling: With Reassignment (P:wR)}
        \label{fig:hist_pool_with_re}
    \end{subfigure}
    \caption{Histograms of differences in actual and communicated waiting time $\Delta^{wt}$.}
    \label{fig:histograms}
\end{figure*}

Table~\ref{tab:eval_scenarios} shows aggregated KPIs of all simulated scenarios. For all scenarios, the service level is similarly high with around 95\% served customers. Compared to the hailing service, even with 400 fewer vehicles, more customers can be served in the pooling case especially in not coupled scenarios. Additionally served customers in the pooling scenarios result in slightly increased Fleet KM compared to hailing. Nevertheless, with shared trips the fleet efficiency measured by average occupancy and saved distance (the relative decrease in Fleet KM compared to all customers driving in private vehicles from origin to destination) increases significantly in the pooling scenarios.

However, once the coupling to Aimsun Next is enabled, all fleet KPIs (except for served customers) change significantly. Especially, the increase of fleet utilization of around 20\% in all scenarios is notable, indicating that the efficiency of the fleet decreases significantly in the presence of dynamic and stochastic travel times. This directly translates to other fleet KPIs of the MoD operator: The fleet vehicle kilometers increase, and saved distance and average occupancy decrease.

The decreased performance can be explained by two effects: 1) Generally, vehicle travel times tend to be longer, which can be observed in the increase in customer travel time in the hailing service from $667$s to $875$s. This effect likely results from the travel time estimation: The average link travel times from the last 30 min are used to predict travel times of the next 30 min. As traffic tends to increase during the simulation period, link travel times are usually underestimated using this method. 2) The stochasticity of travel times leads to infeasible assigned vehicle routes, due to violation of time constraints. The strategy in this paper to deal with infeasibilities is to maintain the assigned schedule until finished. This results in non-optimal assignments, which then results in worse performance as indicated by the KPIs. Allowing re-assignment can improve the service, mainly by mitigating the previous argument. Infeasible schedules are penalized in the objective, and therefore it is tried to assign other vehicles that can fulfill time constraints. 

From customers` perspective, average waiting time and travel times change significantly for all services when the coupling is enabled. Waiting time increases by up to 36\% (P:nR) and travel time by up to 23\% (H:nR). Both effects are likely due to previously discussed underestimation of vehicles travel times.

To measure the reliability of initially communicated pick-up and travel times, the quantities $\Delta^{wt}_{i} = t_{i, pu}^{act} - t_{i, pu}^{exp}$ and $\Delta^{tt}_{i} = t_{i, do}^{act} - t_{i, do}^{exp}$ are defined for each served customer $i$, respectively. $t_{i, pu}^{exp}$ ($t_{i, do}^{exp}$) and $t_{i, pu}^{act}$ ($t_{i, do}^{act}$) correspond to communicated pick-up (drop-off) times with assignment and actual pick-up (drop-off) times, respectively.

Fig.~\ref{fig:histograms} shows histograms of the $\Delta^{wt}$-distributions. All uncoupled scenarios show a very prominent peak at $0$s as deterministic travel times allow a precise forecast of the vehicle arrival. Small deviations can be seen for customer that are picked up when network travel times are updated. Higher deviations can be observed for uncoupled scenarios for pooling with re-assignment because the vehicle to pick up a customer can be different from originally planned and stops for other customers might still be included. For coupled scenarios, the distributions broaden and shift towards positive values, indicating on average delayed pick-ups (between $60$s to $124$s) and less reliable pick-up times (standard deviations of the distribution between $209$s and $339$s). Nevertheless, it can be observed that reassignment can reduce the mean as well as the standard deviation for the $\Delta^{wt}$ distributions from $124$s and $339$s to $60$s and $209$s, respectively. Similar observations can be made for the distributions of the reliability of travel times $\Delta^{tt}$. Compared to $\Delta^{wt}$, this quantity is generally higher for pooling scenarios because new customers can be picked-up on-route resulting in additional travel time.

Computational times exceeded 15h in some pooling scenarios, run on a single Intel Xeon Silver 2.10 GHz processor. Around 75\% of computational is spent on routing queries that cannot be preprocessed in this study.

\section{Conclusion}

MoD services have been an active research topic in recent years. Many studies focused on developing control algorithms to supply efficient services. Most of the algorithms apply hard time constraints to limit the search space, but the effect to deal with dynamic and stochastic travel times which inevitably lead to violations of these constraints are hardly discussed. To evaluate the effects, this study presented a coupled framework of a detailed MoD service simulation with a microscopic traffic simulation. Results showed that the combination of inaccurate travel time estimation and damage control strategies for infeasible routes due to violated time constraints deteriorates the performance of MoD services -- hailing and pooling -- significantly. Also, customers suffer from unreliable pick-up time and travel time estimations. Allowing re-assignments of initial vehicle schedules according to update system states helps to restore system efficiency and reliability, but only to a minor extent.

Overall, this study shows the necessity of further research to 1) the incorporation of more accurate online travel time estimation algorithms that are applicable for an extensive use of routing queries as needed for MoD matching algorithms and 2) more sophisticated damage control algorithms that deal with infeasible routing and assignment solutions due to flawed travel time estimations.

\section{Acknowledgements}

The authors thank the European Union’s Horizon Europe research and innovation programme for providing funding via the project CONDUCTOR under Grant Agreement No 101077049. The authors remain responsible for all findings and opinions presented in the paper.

\bibliographystyle{plainnat}
\bibliography{references}

\end{document}